\documentclass{aa}  

\usepackage{graphicx}
\usepackage{natbib}
\usepackage{txfonts}
\def\kms{\hbox{~km\,s$^{-1}$}\xspace}

\def\msun{~M$_\odot$\xspace}

\def\mjyb{\hbox{~mJy\,beam$^{-1}$}\xspace}

\def\x{$\times$\xspace}
\def\arcsec{$^{''}$\xspace}
\def\deg{$^{\circ}$\xspace}
\def\radec{RA, Dec~(J2000)\xspace}
\def\mic{~$\mu$m\xspace}
\def\co{$^{12}$CO\xspace}

\def\twomass{2MASS\,15430576-3410004\xspace}
\def\iras{IRAS\,15398-3359\xspace}

\usepackage[switch]{lineno}
\begin{document}

   \title{Episodicity in accretion-ejection processes associated with \iras}
   
   \author{E. Guzm\'an Ccolque\inst{1}
          \and
          M. Fern\'andez L\'opez\inst{1}
          \and
          M. M. Vazzano\inst{1}
          \and
          I. de Gregorio\inst{2}
          \and
          A. Plunkett\inst{3}
          \and
          A. Santamar\'ia-Miranda\inst{2}
          }

   \institute{Instituto Argentino de Radioastronom\'ia, CCT-La Plata (CONICET), C.C.5, 1894, Villa Elisa,
Argentina
         \and
        European Southern Observatory, Alonso de Cordova 3107, Casilla 19, Vitacura, Santiago, Chile
        \and   
         National Radio Astronomy Observatory, 520 Edgemont Rd., Charlottesville, VA 22903, USA    }

   \date{Received December XX, 2023; accepted  XX, 2024}

  \abstract
   {The protostar \iras\ is associated with a bipolar molecular outflow ejected in an nearly northeast-southwest (NE-SW) direction which has been extensively studied. It has been suggested previous episodic accretion events by this source. Furthermore, the analysis of the morphology and kinematics of the molecular outflow revealed the presence of four \co(2--1) bipolar elliptical shock-like structures identified in both lobes. These structures seem to trace different ejections inclined $\sim$10\deg on the plane of the sky from each other. This led to the hypothesis that the outflow axis likely precesses and launches material episodically.}
   {Since several authors reached the conclusion of the same episodicity scenario by independent observations, \iras has become an ideal target to empirically analyze the relationship between accretion and ejection processes.}
   {We analyze ALMA archive observations in Band 6, revealing the presence of low-velocity ($<3.5$\kms) emission from the \co(2-1) line to the south and north of the protostar. We study the morphology and kinematics of the gas, which seems to support the hypothesis of a precessing episodic outflow.}
   {The ALMA observations reveal a north-south (N-S) outflow most likely associated with the \iras protostellar system. This outflow could be older than the well-studied NE-SW outflow. The orientation of the N-S outflow is 50\deg-60\deg on the plane of the sky away from that of the NE-SW outflow. We also analyze the Spectral Energy Distribution of a far away young star and preliminary discard it as the driver of the SE outflow remnants. }
   {The new observations support the hypothesis of strong episodic accretion-ejection events in \iras, accompanied by dramatic changes in the orientation of its ejection axis, implying that all the outflows in the region may have been driven by the same protostar.}


   \maketitle
%

\section{Introduction} \label{sec:intro}

Young stellar outflows are believed to extract angular momentum in protostellar disks, allowing material accreting onto the central protostar. However, the link between accretion and ejection is difficult to reach observationally due to the confusion among disk, envelope, outflow and accretion processes, the small scales within the disk, and the inaccuracies of mass estimates. Hence, accretion processes are far from being understood \citep[although see a few empirical work connecting accretion and ejection processes][]{2013Ellerbroek,2023Kim} and, at the same time, many outflows show perturbations moving them away from the traditional view of the well-behaved bipolar outflow \citep{2009Cunningham,Vazzanoetal2021}. 

In this regard, the precession in outflows has usually been related to dynamical interactions of binaries or multiples. However, this effect can be caused in several ways, such as: (1) the orbital motion of a binary system, (2) the tidal effect in the disk by a non-coplanar companion, (3) the warp of the inner disk from which the jet is launched, (4) the misalignment between the spin of the disk and the axis of outflowing ejection \citep[see e. g.][and references therein]{2015Kwon, 2022Young}.
Precession has been reported in outflows from low-mass protostars such as HH30, HH46-47, or L1157 \citep{2007Anglada,Arceetal2013,2015Kwon}. 

Multiple outflows arising apparently from a single young stellar system have been found mostly toward high-mass star-forming regions \citep[e.g., Cepheus~A~HW2][]{2009Cunningham}. The so-called binary jets have also been reported associated with low-mass young stars \citep[e.g.,][]{2008Murphy}, with L1551~IRS5 as a striking example of two jets ejected by two protostars separated by 50\,au \citep[e.g.,][]{2003Rodriguez}. Although the most straightforward explanation for the multiple outflows is to have as many driving objects as bipolar ejections, in certain cases it has been speculated that two outflows can be driven from the same young star \citep{2015Kwon,2009Cunningham}. This can be possible if the outflows are sequentially ejected in time and in different directions. The change of direction could be due, for instance, to the tidal interaction of a non-coplanar companion \citep[as in the case of Cepheus~A~HW2,][]{2009Cunningham,2013Zapata}. Other plausible scenarios could also explain such extreme systems.   

The protostar \iras and its associated molecular outflow has been extensively studied. It is a young low-mass Class\,0 protostellar object located in the Lupus\,I star-forming region at \radec = 15:43:02, --34:09:07. Recent calculations of Lupus\,I distances based on {\it Gaia} DR2 data have revealed that this cloud is located at 153$\pm$5\,pc \citep{Santamaria-mirandaetal2021}, in agreement with the distance derived by \citet{Sanchisetal2020}.

\iras was first identified by \citet{HeyerandGraham1989}. Its associated outflow was first reported by \citet{Tachiharaetal1996} and mapped in several CO transitions by \citet{vanKempenetal2009b} via single-dish observations.

From high angular resolution observations of H$_2$CO and CCH obtained with ALMA, \citet{Oyaetal2014} detected the molecular outflow extending in the northeast-southwest direction (PA 220\deg), and derived an inclination angle of 20\deg with respect to the plane of the sky. They estimated an upper limit of 0.09\msun for the protostellar mass.
Using SO ALMA observations, \citet{2018Okoda} suggested the presence of a molecular gas disk, which has been recently resolved out by \citet{2023Thieme} into a 31.2~au radius structure. In this last work, the authors dynamically derived a protostellar mass of 0.022\msun, in good agreement with that estimated by \citet{Yenetal2017} and \citet{2018Okoda}. This lower mass value makes IRAS~15398-3359 an object between the proto-brown dwarf and the very low-mass regime. 
Furthermore, the reported envelope mass ranges from 0.5 \citep{vanKempenetal2009} to 1.2\msun \citep{Jorgensenetal2013}, suggesting the protostellar growth in the future.

\citet{Jorgensenetal2013} and \citet{Bjerkelietal2016b} have suggested previous episodic accretion events by this source. \citet{Jorgensenetal2013} detected a H$^{13}$CO$^+$ ring structure of about 150-200\,au around the protostar. The lack of H$^{13}$CO$^+$ inside the ring is not consistent with the current heating rate of the central protostar. These authors propose that the H$^{13}$CO$^+$ would have been removed by a chemical reaction with H$_2$O, sublimated from dust grains during an accretion burst that occurred 10$^2$--10$^3$ years ago.
\citet{Bjerkelietal2016b} also provided evidence for a past accretion event via the study of HDO(1$_{0,1}$--0$_{0,0}$). 
The authors found this molecule is only detected in the region closest to the protostar, and they suggest as a possible explanation that the water in the grains was released during a recent accretion burst. Furthermore, the analysis of the morphology and kinematics of the northeast-southwest molecular outflow revealed the presence of four pairs of counter-aligning elliptical shock-like structures identified in both lobes \citep{Vazzanoetal2021}. These structures seem to trace different ejections inclined $\sim$10\deg from each other. This led to the hypothesis that the outflow axis likely precesses and launches material episodically. Since several authors reached the conclusion of the same episodicity scenario by independent observations \citep{Jorgensenetal2013,Bjerkelietal2016b,Vazzanoetal2021}, \iras has become an ideal target to empirically analyze the relationship between accretion and ejection processes.

From ALMA 12m array observations in Band 6, \citet{Okodaetal2021} detected the arc-like structure in H$_2$CO, SiO, CH$_3$OH and SO crossing the northeast-southwest  molecular outflow in a direction roughly perpendicular. They proposed that the observed feature could be part of a relic outflow ejection, previously launched by \iras. They interpret the difference in the ejection direction of this relic with respect to the direction of the northeast-southwest outflow as induced by variations in the angular momentum of the episodically accreting gas. These variations could produce a drastic change in the direction of outflow ejection. Alternatively, \citet{Vazzanoetal2021} reveal a complex of \co(2--1) and SO (J$_N$=6$_5$, 5$_4$) arc-like structures, 10\arcsec--20\arcsec southeast of the protostar's location. Some of these structures present bow-shock shapes with tips pointing toward \iras, suggesting a possible origin linked to the emission of an outflow associated with the source \twomass, placed about 1$\arcmin$ southeast of \iras, \citet{Okodaetal2021} interpreted these arc-like features as coming out from a possible second outflow associated with \iras.
In the present contribution we analyze recent ALMA archive observations, revealing the presence of red-shifted and blue-shifted low-velocity gas detected north and south of \iras, respectively. This confirms the existence of a other outflow driven by this protostellar system. We study the morphology and kinematics of the gas, which seems to support the hypothesis of a precessing episodic outflow.

\section{ALMA molecular data}
This work is mainly based on archival data obtained with the 7m Atacama Compact Array (ACA) of ALMA. We also use higher angular resolution images obtained with ALMA, previously presented in \citet{Vazzanoetal2021}. These images were not combined and we analyzed them independently. New archival data presented in this work consist of a mosaic done with the ACA 7m-antennas array in Band 6  centered on IRAS 15398-3359 (project 2019.1.01063.S, P.I: Jinshi Sai). The mosaic contained 28 pointings covering an area of 2\farcm4\x2\farcm6. The data were taken on December 19, 2019. Maximun and minimum baselines were 8 and 48 meters, then the angular resolution is 8\arcsec. With respect  to  the  weather  conditions,  the precipitable  water vapour (PWV) values ranged between 1.6 and 1.95\,mm during the observations. The correlator was configured to use four spectral windows and one continuum window. The continuous window at 1.3\,mm (233.999\,Ghz) has 128 channels and a bandwidth of 2.0\,Ghz. The spectral windows were centered on the transitions of  C$^{18}$O~(2-1), $^{13}$CO~(2-1), CO~(2-1) and N$_{2}$D$^{+}$~(3-2) at 219.561, 220.399, 230.539 and 231.323 Ghz, respectively. The number of channels and the bandwidth of the first two windows were 2048 and 0.125 GHz each, while for the remaining two windows, 1024 channels were observed with a bandwidth of 0.062\,GHz (equivalent to a velocity resolution of 0.08\,\kms). Calibration of the raw visibility data was performed using the standard reduction script for the Cycle\,6 data provided by the ALMA Observatory. This pipeline ran within the Common Astronomical Software Application \citep[CASA\,5.6.1][]{McMullin2007} environment. The on source integration time was 2.55 hours and the calibrators used to correct for instrumental and atmospheric disturbances (flux, phase, and bandpass) were J1337-1257, J1534-3526, and J1337-1257 respectively. The self-calibrated interferometric free-line continuum data were cleaned in CASA to produce continuum images. The spectral line cubes were produced  by  subtracting  the  continuum and  applying  a  standard cleaning with primary beam correction. 
The continuum was subtracted in the uv-plane using the \texttt{uvcontsub} task.
The ACA calibrated visibilities were Fourier transformed and cleaned with the CASA task \texttt{tclean}. We set the Briggs weighting parameter robust$=0.5$ for both the continuum and \co(2-1) images as a compromise between angular resolution and signal-to-noise ratio (beam of 7\farcs7$\times$4\farcs2, PA=86\deg). The rms noise level in the continuum image is around 2\,mJy/beam. The rms noise level for the \co(2-1) cube is $\sim$100\,mJy~beam$^{-1}$ per one channel of 0.16\,\kms of the line velocity cube. In this work we only report \co(2-1) data.


\section{Results} \label{sec:results}

\begin{figure}[!t]
  \includegraphics[width=0.5\textwidth]{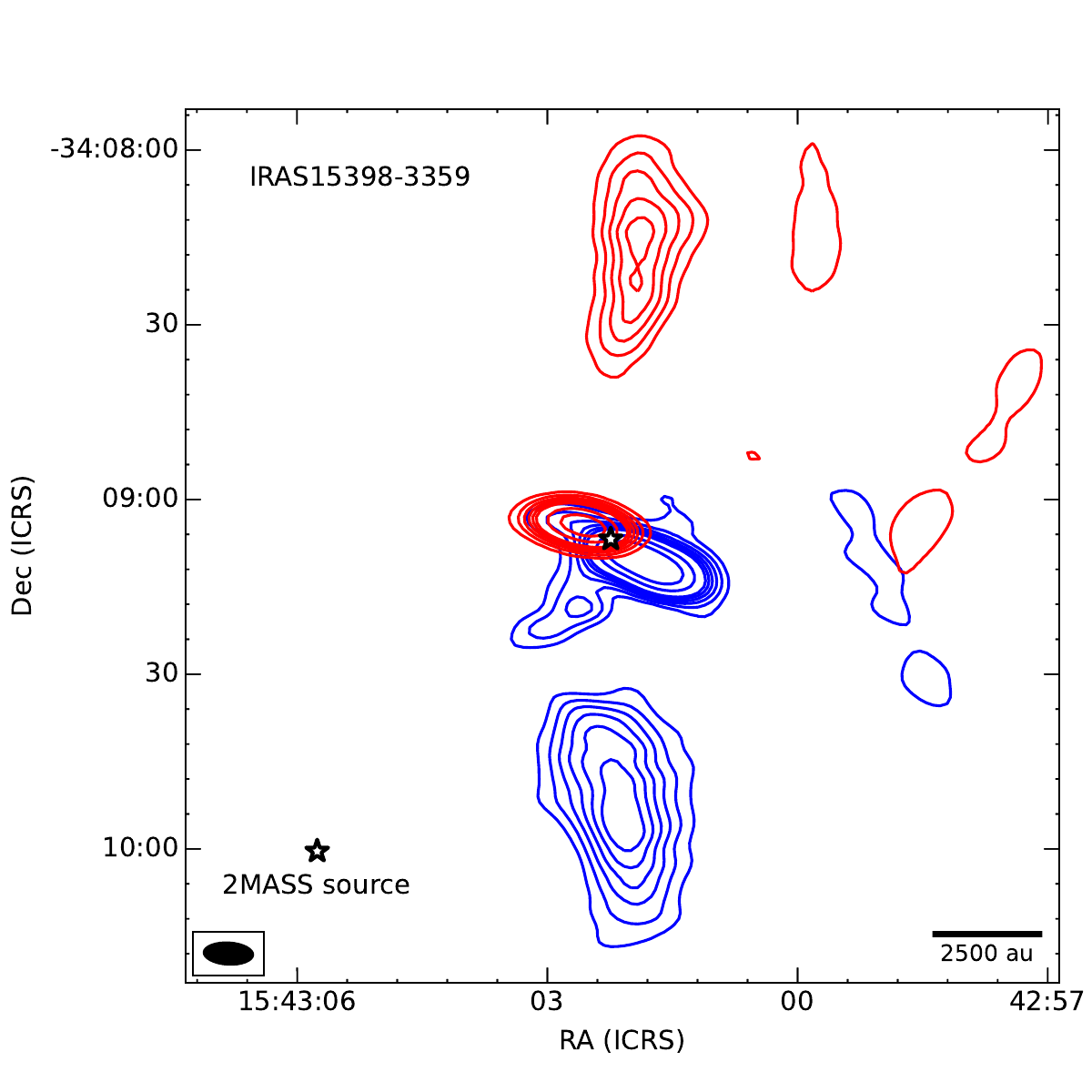}
  \caption{Blue-shifted and red-shifted integrated emission \textbf{of \co(2-1)} toward the \iras region. The blue-shifted and red-shifted emission were integrated over the velocity ranges from 2.3 to 4.7~\kms and from 6.2 to 7.4~\kms. Blue-shifted  and red-shifted contours are represented at 10, 20, 30, 40, 50, 60, 100, and 150 times the $rms$ of 0.15\,\mjyb\kms and 0.1\,\mjyb\kms, respectively. The stars indicate the position of \iras and \twomass. The synthesized beam (7\farcs7$\times$4\farcs2, PA=86\deg) is represented by the black ellipse in the bottom left corner.}
  \label{fig:moment0}
\end{figure}

Fig. \ref{fig:moment0} shows the blue-shifted and red-shifted integrated \co(2-1) intensity images, considering a systemic velocity v$_{sys}$= 5.1$\pm$0.1~\kms \citep{Mardonesetal1997,vanKempenetal2009b}. This systemic velocity is consistent with the recent ALMA study \citep{Yenetal2017}.
In the center of the figure there is strong emission corresponding to the well-known molecular flow associated with the northeast-southwest direction. Moreover, there is a second lobe to the southeast, which was identified as an outflow relic by \citet{Okodaetal2021}.
In addition, we detected other blue-shifted and red-shifted gas emission to the south and north of \iras, respectively. The morphology seems an additional molecular outflow. These structures of \co are aligned with respect to the location of the \iras protostar and their emission is fainter than that of the northeast-southwest outflow. In addition, their projected size in the plane of the sky is greater than the northeast-southwest outflow. Hereafter, we further refer to the well-known outflow in the northeast-southeast direction as the `NE-SW outflow', the southeast lobe reported by \citet{Okodaetal2021} as the `SE lobe', and to the new one detected in the north-south direction as the `N-S outflow'. 

Table \ref{tab:param_outflow} lists the velocity ranges used to map the red-shifted and blue-shifted lobes of the NE-SW and N-S structures, the position angles (P.A.) measured from north (0\deg) to east (90\deg), and the size of lobes. We derive these quantities for the red- and blue-shifted lobes of each outflow separately. $\Delta v_{rad} =|v_{sys}-v_{max}|$ is the outflow spread in radial velocity, where $v_{sys}$ is the systemic velocity and $v_{max}$ is the maximum velocity which it is possible to detect outflow emission over the 3$\sigma$ threshold. The position angles are calculated from the source position to the peak emission in each lobe.
Since the outflow inclination is unknown, the sizes are projected on the plane of the sky and should be treated as lower limits.  The errors in the measured sizes are given by the angular resolution of the image ($\Delta$Size = beam/2).  The parameters listed for the NE-SW outflow were taken from \citet{Vazzanoetal2021}.

\begin{table}
    \centering
    \begin{tabular}{c|ccc}
        \hline
             & $\Delta$v$_{rad}$ &   PA   & Size  \\
             &       (\kms)       & (\deg)  & (au)  \\  
                \hline 
        \textit{NS outflow} &  &  & \\
        \hline
        Blue lobe &  3.4$\pm$0.08   & 182$\pm$4.0 & 10030$\pm$570 \\
        Red  lobe &  2.9$\pm$0.08   & 355$\pm$4.0  & 10100$\pm$570 \\  \hline              
        \textit{NE-SW outflow} &  &  &    \\
        \hline
        Blue lobe &  15.3$\pm$0.08  & 232.0$\pm$0.2 & 2550$\pm$50 \\
        Red  lobe &  11.6$\pm$0.08  &  64.9$\pm$0.2 & 1800$\pm$50 \\   \hline 
 
 
        \hline
    \end{tabular}
    \caption{Outflow parameters. $\Delta v_{rad} = |v_{sys}-v_{max}|$ is the outflow spread in radial velocity, where $v_{sys}$ is the systemic velocity and $v_{max}$ is the velocity up to which it is possible to detect outflow emission over the 3$\sigma$ threshold.}
    \label{tab:param_outflow}
\end{table}

\begin{figure}[!t]
\centering
  \includegraphics[width=1.0\columnwidth]{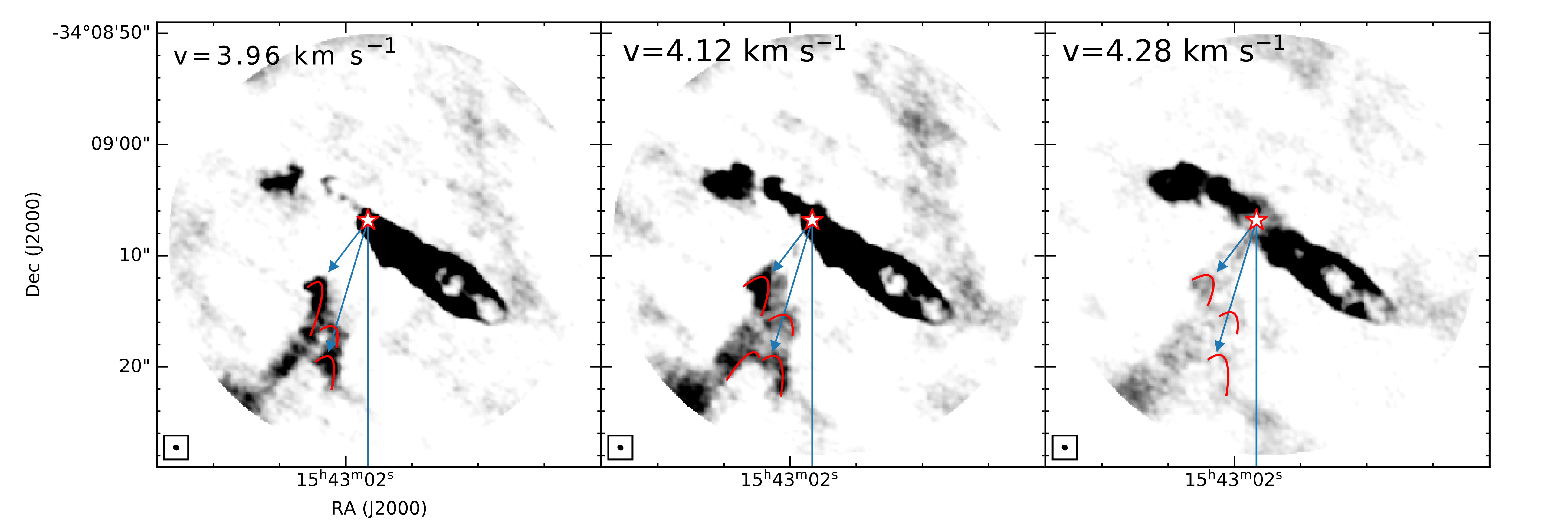}
  \includegraphics[width=0.69\columnwidth]{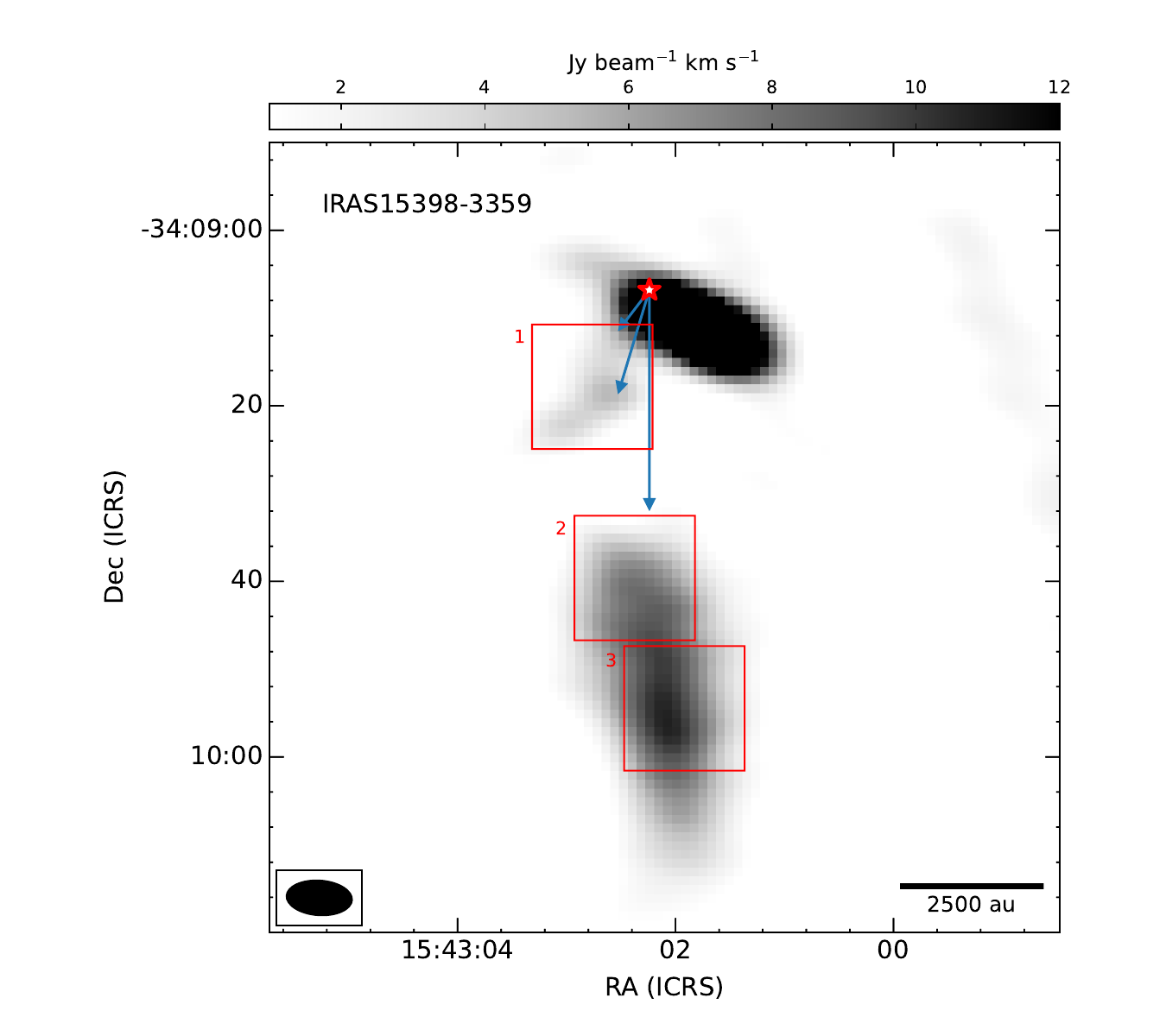}
  \includegraphics[width=0.30\columnwidth]{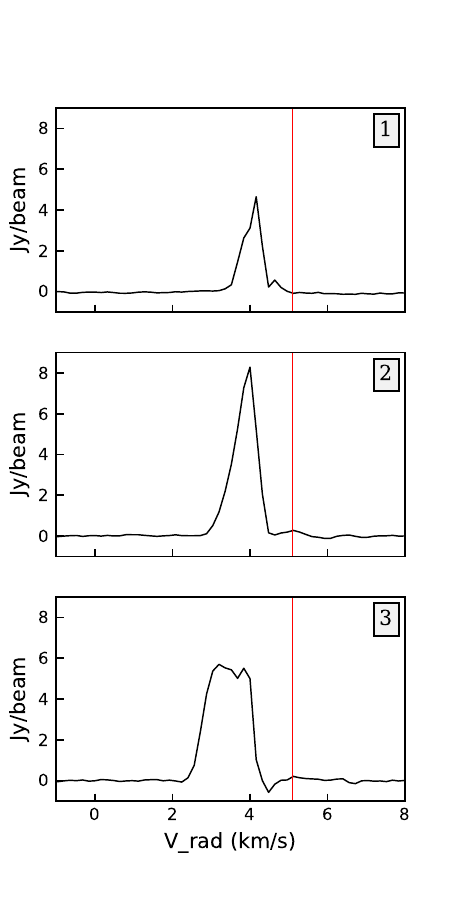}
  \caption{\textit{Upper panels}: High-angular resolution velocity channel maps taken with the 12-m array of the \co(2-1) emission near the cloud velocity. The blue arrows indicate possible relic ejections interacting with the gas in the vicinity of \iras. The red star indicates the position of the continuum source. The synthesized beam (0\farcs57$\times$0\farcs50) is represented in the bottom left corner and the radial velocity is indicated in the top left corner. \textit{Bottom left panel}: The same arrows as in the figure above are shown on the integrated blue-shifted emission image covering from 2.3 to 4.7~\kms. The red boxes show the regions within which the spectra shown on the right-hand panels were taken. \textit{Bottom right panels}: Average spectra obtained integrating the blue-shifted emission inside the three boxes from the left. The vertical red line  in the spectra indicates the systemic velocity at 5.1\,\kms.}
  \label{fig:ejections}
\end{figure}

In the top panels of Figure\,\ref{fig:ejections} we show three channels of the high-angular resolution \co\,(2-1) velocity cube \citep{Vazzanoetal2021}. The emission in these channels is slightly blue-shifted with respect to the system velocity. The NE-SW outflow is very bright, and a fainter complex of several arc-shaped structures (that we have identified as the SE lobe) can be identified southeast of the protostar location, almost perpendicular to the NE-SW outflow. These arc-shaped structures are also detected in CO(2-1) by \citet{Vazzanoetal2021}, and other molecular tracers (H$_{2}$CO, SO, SiO and CH$_{3}$OH) in \citet{Okodaetal2021}. 
We identify two main directions linking the protostar position with the tips of some prominent arc-shaped structures; we include a third direction pointing to the blue-shifted southern lobe of the N-S outflow. These three directions, marked with blue arrows in Fig.\,\ref{fig:ejections}, have position angles 137\deg, 160\deg\ and 180\deg. They seem to be separated by $\sim$20\deg each on the plane of the sky. The three panels in the bottom right of Fig.\,\ref{fig:ejections} present the mean integrated spectra taken toward three regions marked with the corresponding boxes in the bottom left panel. Spectra from boxes 1 and 2 show a single main peak blue-shifted at 4.1\,\kms. The spectrum from box 3 shows a double peak structure at 3.8\,\kms and 3.2\,\kms.

\begin{figure}[!t]
  \includegraphics[width=1\columnwidth]{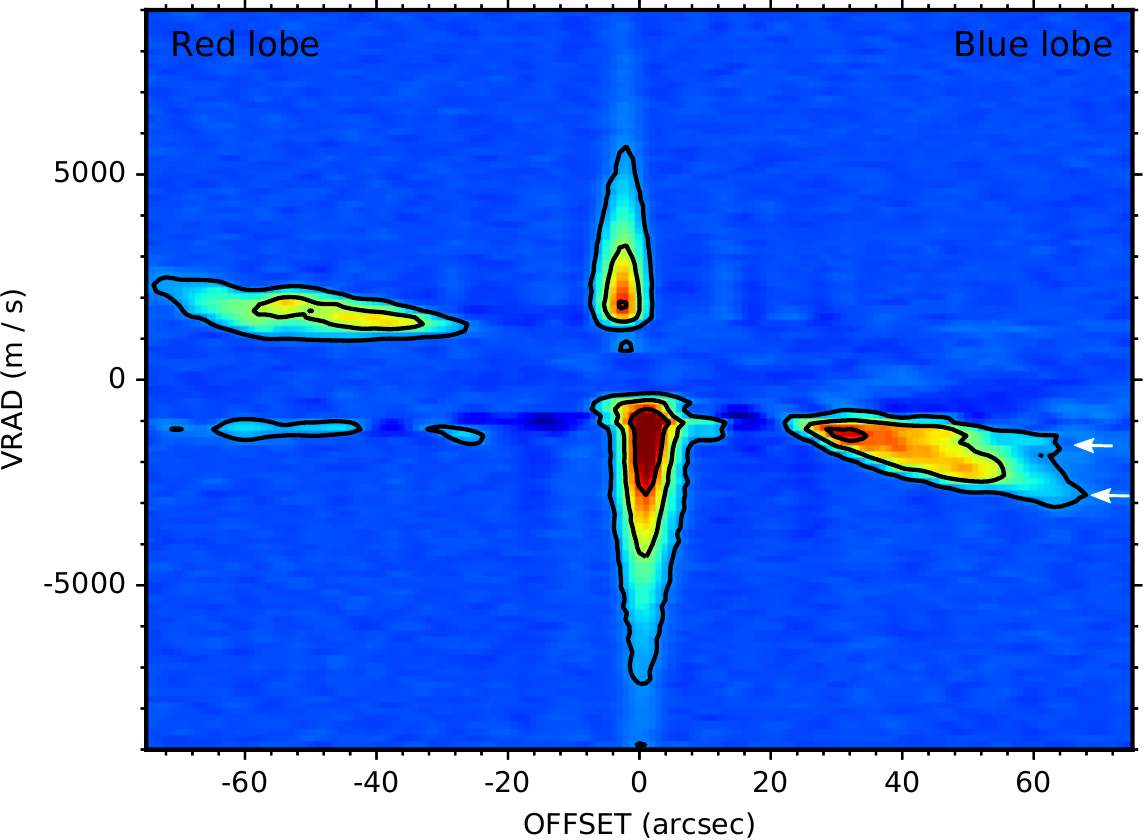}
  \caption{Position-velocity diagram of the \co(2-1) emission along the north-south axis and centered at the continuum peak position with an angle position of and cut width of 1\arcsec.}
  \label{fig:pv}
\end{figure}

Figure \ref{fig:pv} shows the position-velocity diagram obtained along the N-S outflow axis and centered at the continuum peak position and the systemic velocity (5.1~\kms). The emission near the zero offset  likely belongs to the \iras\ rotating molecular envelope/disk system. Further away, the red-shifted and blue-shifted gas moves up to 70\arcsec\ away from the protostar and reaches velocities as fast as 3\,\kms\ with respect to the systemic velocity.
The two-branch morphology observed in the blue-shifted part of the diagram (positive offsets) could correspond to a shell structure typical of molecular outflows (see white arrows in the Figure\,\ref{fig:pv}). This agrees with the double-peaked spectrum corresponding with box 3 in Figure\,\ref{fig:ejections}.

\begin{figure}
    \centering
    \includegraphics[width=0.5\textwidth]{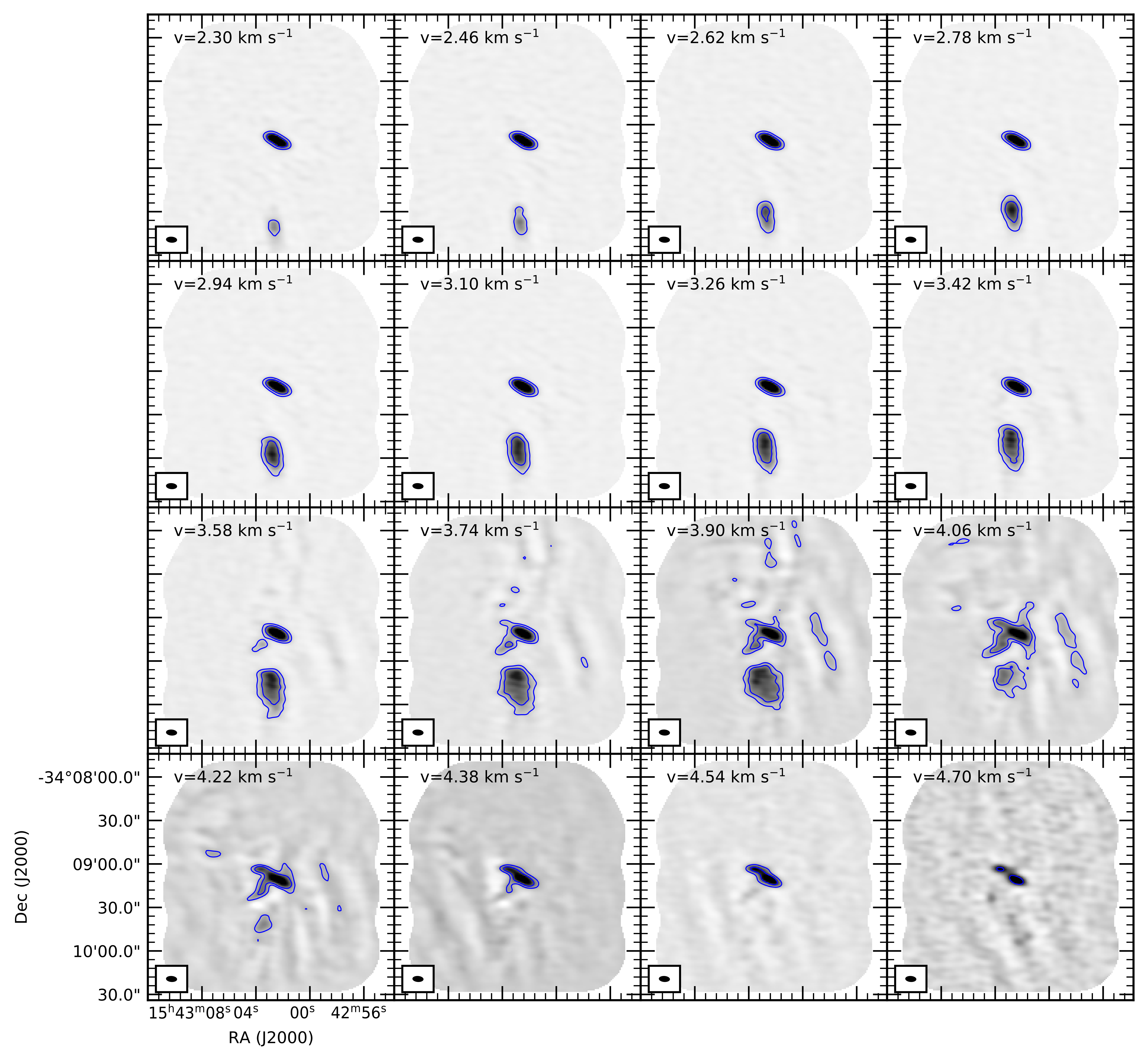}
    \includegraphics[width=0.5\textwidth]{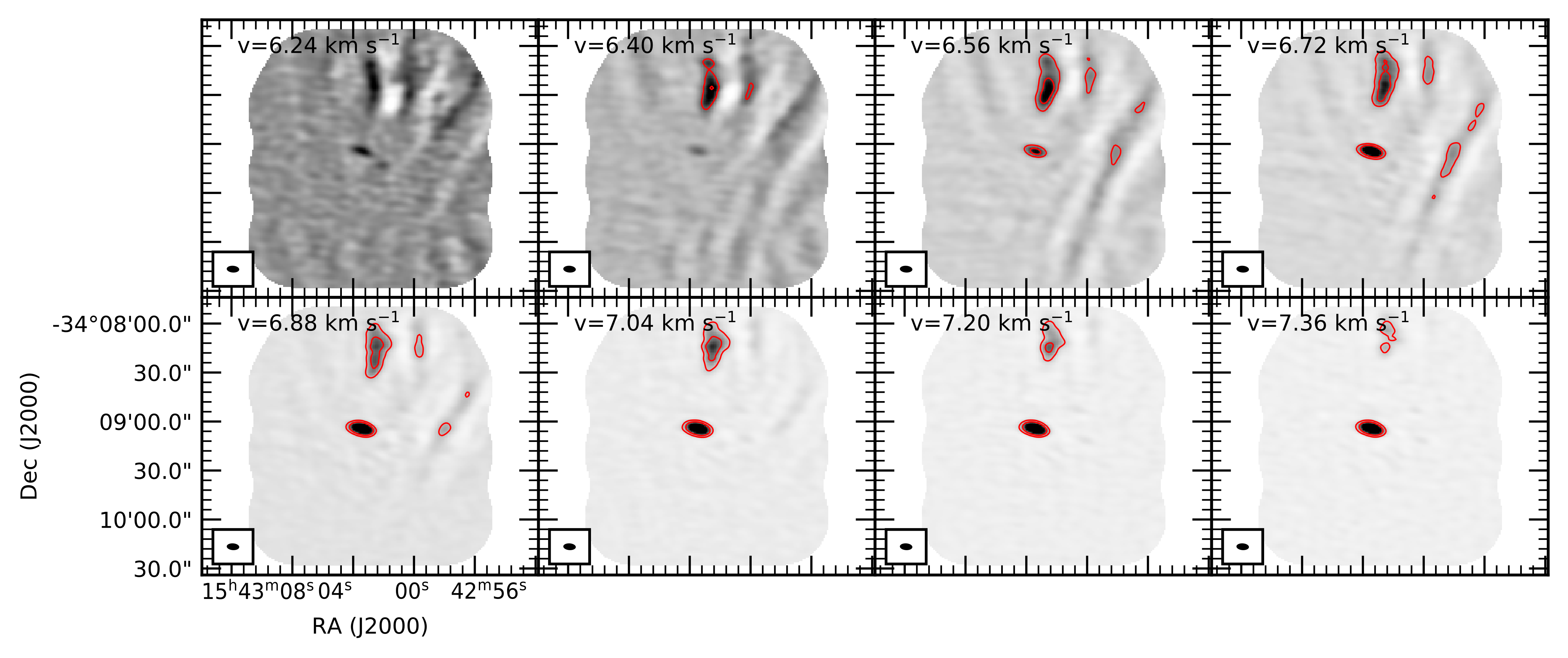}
    \caption{\co emission velocity channel maps toward \iras (grey scale). The upper/lower channel map shows the blue/red-shifted emission, avoiding the central channels around the 5.1\kms cloud velocity. Contours are displayed at 2.0 and 5.0\,\mjyb. Radial velocities are indicated in the top left corner. Synthesized beam is shown in the bottom left corner on every channel.}
    \label{fig:chan}
\end{figure}

Figure\,\ref{fig:chan} shows the $^{12}$CO (2-1) velocity channel map of the NE-SW and NS outflows, and the SE lobe. The velocity of the blue-shifted emission from the NS outflow ranges from 2.3 to 4.2\kms and its red-shifted emission ranges from 6.4 to 7.4\kms. The SE lobe emission ranges from 3.6 to 4.4\kms. The NE-SW main outflow spreads a range larger than that covered in the Figure.

From the \co map in Fig.\,\ref{fig:moment0} and the velocity gradients of the gas revealed in Fig.\,\ref{fig:pv}, we can infer the presence of two bipolar structures centred on the IRAS 15398-3359.  The NE-SW outflow that has been extensively studied and the new N-S outflow presented in this work for the first time.  A multiple outflow scenario supports the scenario proposed by Okoda et al. (2021), who revealed various outflows coming out from IRAS15398-3359 at different epochs (\ref{sec:discussion}). The position angles of the blue-shifted and red-shifted lobes of the N-S and NE-SW outflows differ by 173\deg and165\deg, respectively. In addition, the size projected on the plane of the sky of the N-S outflow is over four times greater compared with the NE-SW outflow, while its velocity extent is at least four times smaller without correcting for inclination angles.
\section{Discussion} \label{sec:discussion}
\subsection{A north-south molecular outflow from \iras}

In addition to the well-known NE-SW molecular outflow associated with the protostar \iras, \citet{Okodaetal2021} showed indications of the presence of a second molecular outflow probably ejected by the protostellar system, extending southeast up to 2\arcsec (3000\,au). The present 7m\,array ALMA observations reveals the existence of a third molecular bipolar outflow in the north-south direction. The projected sizes measured from the position of the protostellar system ($\sim10^4$\,au) and the radial velocities (up to $\sim3$\kms) of the blue-shifted (south) and red-shifted (north) lobes are similar; their position angles match within the error bars. All of this supports the idea that these two lobes comprise a bipolar structure originated at the \iras\ position.
Similar cases can be found in the literature where more than one outflow is observed. There is the case of Par-lup where two outflows were found associated with a young very low mass star \citep{2020Santamaria}. In this source the evidence points to a $<15$\,au packed binary driving the two outflows. Another example is given in L1157, where \cite{2015Kwon} observed two jets would be of different ages and would be triggered by a single Class 0 protostar ($<0.04$\msun).

\subsection{Origin and nature of SE lobe}
Furthermore the 7m\,array ALMA observations show, although unresolved, the emission from the blue-shifted complex of arc-shaped structures (the SE lobe), previously detected by \cite{Okodaetal2021} in several molecular shock-tracers between 1000\,au and 3000\,au southeast of the protostar (position angles ranging 137\deg-160\deg). The red-shifted counterpart of these arc-shaped structures is, however, not clearly seen with the current low-angular resolution data \citep[but see the $^{13}$CO and C$^{18}$O velocity cubes in ][]{2023Thieme}. In the following, we discuss the origin and nature of the molecular emission of the SE lobe and its relationship with the two outflows apparently launched from \iras location.

\begin{figure}
    \centering
    \includegraphics[width=0.5\textwidth]{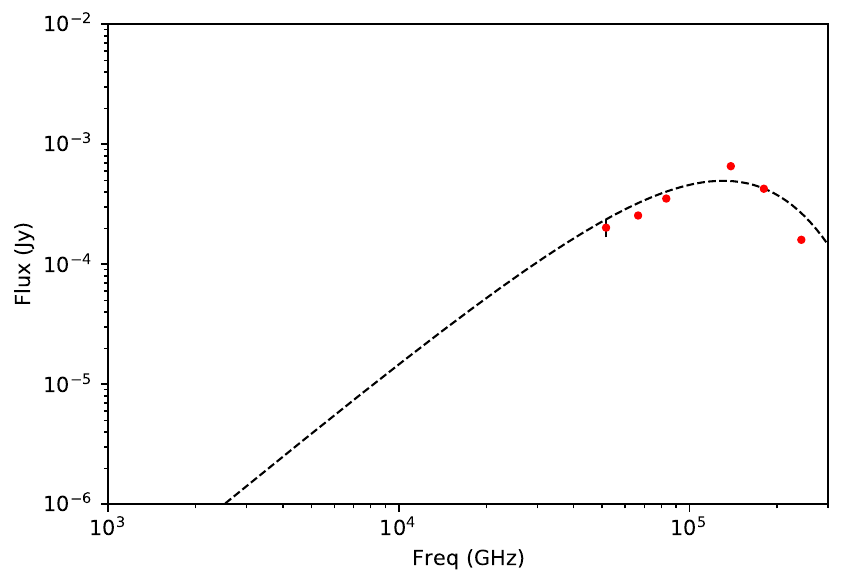}
    \caption{\twomass\ spectral energy distribution. Black dashed line shows the black-body fitting derived from infrared emission (data listed in Table\,\ref{tab:twomass}).}
    \label{fig:sed}
\end{figure}

Regarding \iras, a first possible scenario could be the presence of another source to the southeast, launching an outflow that produces bow-shock arc-shaped structures when breaking into the quiescent gas surrounding \iras \citep{Vazzanoetal2021}. Figure\,\ref{fig:moment0} shows the position of the southeastern infrared source \twomass, about 9000\,au away ($\sim60''$). However, the molecular gas from the arc-complex does not extend to this location. In addition, we have compiled and analyzed the SED of this 2MASS source. The SED can be fitted with a blackbody model of temperature 2200\,K, which peaks at $\sim$10$^5$\,GHz (Figure\,\ref{fig:sed}; see also Table\,\ref{tab:twomass}). The infrared spectral index from 2 to 24\,$\mu$m is $\alpha$=$\frac{dlog(\lambda F(\lambda))}{dlog\lambda}$ = --2.27. The lack of strong millimeter emission, along with the derived infrared spectral index, indicates that this star is probably in a more evolved stage \citep{Evans2003}, and is not probably responsible for the ejection of a prominent outflow. 

A second possible scenario, proposed by \citet{Okodaetal2021}, suggests that the arc-complex originates from relic \iras\ ejections. In this scenario, the geometry of the arcs is showing the bottom part of bubble-shaped structures, instead of bow-shocks tips.  A variation of the SE lobe as a relic outflow scenario, would explain the arcs and the filamentary structure (upper panels in Figure 2) as the remnants of a side-way shock created by adjacent wakes of two separate ejections. This may explain the arcs pointing toward the protostar and the shock tracers found in this structure. Ejections with this type of morphology (cavities with parabolic shapes) are considered in the wind-driven shell model explained by \citealt{Leeetal2000}, and have already been detected in other outflows, such the northwest lobe of IRAS\,16059-3857 \citep{Vazzanoetal2021} or the western lobe of HH\,46/47 \citep{Arceetal2013}. Alternatively, the SE lobe could comprise the relics of two colliding side-way shocks from adjacent outflow ejections. The discovery of N-S outflow extending up to $10^4$\,au supports this second hypothetical scenario, as it provides new evidence for the existence of older ejecta launched from the \iras location.

The difference in the position angles of the different arcs and the southernmost blue-shifted lobe ($\sim$20\deg-23\deg) may indicate that the outflow axis was precessing. A similar set of ejections in slightly different directions have also been observed in the NE-SW outflow \citep{Vazzanoetal2021}. We analyze this precessing behaviour in the Section \ref{sec:episodic}.

\begin{table}
    \centering
    \resizebox{9cm}{!}{
    \begin{tabular}{cccc}
    \hline
    Band & Wavelength & Flux  & Reference \\
         &   [\mic]   & [mJy] &  \\
    \hline
    2MASS J       & 1.235 &  0.160$\pm$0.033 & \citet{Cutrietal2003}  \\
    2MASS H       & 1.662 &  0.426$\pm$0.058 & \citet{Cutrietal2003} \\
    2MASS K$_s$   & 2.159 &  0.657$\pm$0.065 & \citet{Cutrietal2003} \\
    Spitzer-IRAC1 & 3.6   &  0.353$\pm$0.020   & Evans et al. (2003) \\
    Spitzer-IRAC2 & 4.5   &  0.255$\pm$0.016   & Evans et al. (2003)\\
    Spitzer-IRAC3 & 5.8   &  0.202$\pm$0.034   & Evans et al. (2003)\\
    \hline
    \end{tabular}
    }
    \caption{Data used to fit the \twomass\ spectral energy distribution.} 
    \label{tab:twomass}
\end{table}

\subsection{The age of the relic N-S outflow}
The inclination of the outflows with respect to the plane of sky is still undetermined. Given that the NE-SW outflow seems to change its direction in short time lapses, the inclination with respect to the plane of the sky derived may be different to the value at which the relic N-S outflow was ejected. Moreover, their dynamic times depend on the observed radial velocities, which are considerably lower in the N-S outflow than in the NE-SW outflow. Although the velocities of the outflows are not a robust indicator of their age, this could mean that the N-S outflow was ejected in a direction closer to the plane of the sky (i.e., $i\sim0$\deg), and/or that the gas has slowed down after some time. In any case, older outflows are expected to be longer and slower than younger ones which hints at a difference in the age of both outflows.

As a first approximation, we consider the inclination of both outflows to be the same \citep[$i=30$\deg, from][]{Yenetal2017}. We also use the radial velocities listed in Table\,\ref{tab:param_outflow}. The dynamical times for the blue-shifted and red-shifted lobes of the N-S outflow result in 24200 and 28600 years, respectively. This implies that the N-S outflow would be about 20 times older than the NE-SW outflow (with a dynamic time of $\sim1300$\,years). 
Alternatively, we could also speculate that the gas in the N-S outflow was ejected in the past at similar velocity to that observed in the NE-SW outflow (i.e., a radial velocity of 13.5\,\kms on average). Under this hypothesis, the N-S outflow would have an inclination of $\sim13$\deg with respect to the plane of the sky, and an estimated dynamical time of $\sim$3650\,yr, about 4 times older than the NE-SW outflow.

To sum this up, the new observations suggest that the N-S may be a relic of an older outflow, while the NE-SW outflow, currently being ejected, may be younger.

\subsection{Episodic accretion and ejection}\label{sec:episodic}

\citet{Vazzanoetal2021} identified four pairs of bipolar elliptical structures ending in bow-like structures in the NE-SW outflow associated with \iras. Every identified ellipse-like structure would correspond to a bipolar ejection. These structures have slightly different sizes and position angles, which possibly indicates the presence of episodically ejected material outflowing from the protostellar system with a variable direction caused by the precession of the launching axis.
Extending this scenario to the SE lobe,  we speculate that both, the N-S outflow and the SE lobe, can be part of a series of ejections in slightly different directions with position angles separated by $\sim20$\deg\ (see arrows in Fig.\,\ref{fig:ejections}). Therefore, the NE-SW outflow on one hand, and the N-S outflow along with the SE lobe, on the other hand, may share a similar pattern of episodic precessing ejections.

Assuming that the NE-SW outflow, N-S outflow and SE lobe are driven by the same protostellar system in \iras\ \citep[for an update about the multiplicity of this system based on the most recent observations see][]{Okodaetal2021,2023Thieme}, the present observations suggest that events would have drastically changed the outflow direction, while keeping memory of the precession of the system (or as a consequence of it). In this way, the system would present both, small ($<20$\deg) and large ($>20$\deg) orientation changes. While the small changes may be caused by precession of the outflow axis, the large changes may be triggered by cataclysmic accretion events (a sudden asymmetric accretion of large amounts of mass), the close encounter with an interloper, or by the presence of an undetected small-mass companion star with a nearby periastron \citep[reorientation by gravitational tugging, as in Cepheus\,A\,HW2,][]{cunninghametal2009}. Another possibility is that the gravitational interactions of a putative unresolved multiple system may episodically tilt the system, with intertwined periods of quiescent and chaotic reconfigurations. However, as said before, at this moment there is no clear evidence supporting the presence of such multiple system. Recent continuous analysis of high-resolution ($\sim40\,mas$) \iras observations by \cite{2023Thieme} has revealed one small, compact (deconvolved size of $4.5\times2.8\,au$) and very low-mass ($0.6-1.8\,M_{jup}$) dust disk.

In any case, the data reveal that the \iras\ system has different episodic ejections, and these could be associated with events in which the accretion suddenly varies. Moreover, \citet{Vazzanoetal2021} determined that the estimated dynamical times of the different ejections in the NE-SW outflow range from 33 to 268 years, while \citet{Jorgensenetal2013}, based on the lack of H$^{13}$CO$^+$ at the center of the protostellar system's envelope, proposed that the system underwent a recent accretion burst 100-1000 years ago. Despite the rough agreement between the two timescales, more data should be collected to relate this accretion event  with the sudden change of direction in the outflow.

Mosaic observations with better angular resolution, higher sensitivity (e.g., 12\,m ALMA observations with a extended array) will be necessary to accurately describe the structure of the relic N-S outflow lobes and discover if there are more continuum sources in IRAS 15398-3359. It would also be necessary to observe new chemical tracers to study the accretion processes of the protostellar envelope. In addition, observations with a larger field, including the whole extension of the SE lobe, could help us to better determine its origin.
\section{Summary}

From 7m array ALMA data we report the presence of a new molecular outflow (the N-S outflow) associated with the young protostar \iras, in addition to the already well-studied one projected into the sky in a northeast-southwest direction. The NE-SW outflow may undergo precessional motions, which result in ejections driven along slightly $\sim10$\deg different orientations \citep{Vazzanoetal2021}. The newly reported molecular outflow is almost in a north-south direction and its position angle differs by 50\deg-60\deg\ from the NE-SW outflow. The morphology and kinematics of the detected gas show that the north-south ejections are older and may be the relics of a past ejected bipolar outflow. In addition, we build up the Spectral Energy Distribution of the 2MASS\,15430576/3410004, which shows a spectral index that momentarily discard it as the possible driver of the outflow remnants in the form of arc-shaped structures, previously detected southeast from \iras \citep{Cutrietal2003,Evans2003}. These remnants may be the outcome of the collision of wakes produced by side-way shock fronts.

We propose that the gas detected north and south of \iras\ in these new observations, together with the gas structures identified southeast of the same source \citep{Vazzanoetal2021,Okodaetal2021}, are the remnants of several episodic outflow ejections driven by a possible precessing system.

The new observations suggest that one or more events could have drastically changed the direction of the outflow. This result supports the scenario proposed by \cite{Okodaetal2021}. They propose strong events could be related with extreme anisotropic accretion events that may produce a tilt in the direction of the disk-protostar system. In the case of \iras, both the disk and the protostar have a relatively low mass, hence this sort of perturbation may not be so difficult to reach. After the tilt, or as a consequence of it, the system seems to keep the memory of such events in the form of a small precession seen as multiple ejections in the NE-SW outflow. It would be expected that the rotation of the disk may dampen the precession in due time.

Finally, the results presented in this work, depict \iras  as a key protostellar system to better understand the link between accretion and ejection, their episodic nature, and the origin of precession in very young jets and outflows.

\section*{Acknowledgements}
This paper makes use of the following ALMA data:
ADS/JAO.ALMA 2019.1.01063.S and and ADS/JAO.ALMA 2013.1.00879.S. ALMA is a partnership of ESO (representing its member states), NSF (USA) and NINS (Japan), together with NRC (Canada), MOST and ASIAA (Taiwan), and KASI (Republic of Korea), in cooperation with the Republic of Chile. The Joint ALMA Observatory is operated by ESO, AUI/NRAO and NAOJ. The National Radio Astronomy Observatory is a facility of the National Science Foundation operated under cooperative agreement by Associated Universities, Inc. 
E.G.C. is funded by PhD scholarships from the Argentine Consejo Nacional de Investigaciones Cient\'ificas y T\'ecnicas (CONICET).

\bibliography{biblio.bbl}
\bibliographystyle{aa}

\end{document}